
\input harvmac

\Title{\vbox{\baselineskip12pt\hbox{USC-93/021}}}
{\vbox{\centerline{The winding angle distribution }
 \vskip4pt\centerline{for Brownian and
 SAW revisited}}}

\centerline{Hubert Saleur\foot{Packard Fellow}}
\bigskip\centerline{Physics Dept. and Math Dept.}
\centerline{University of Southern California}
\centerline{Los Angeles CA 90089-0484, USA}
\vskip.3in
We study some generic aspects of the winding angle
 distribution around a point in two dimensions
for Brownian and self avoiding walks
(SAW)  using corner transfer matrix
and conformal field theory.
\newsec{Introduction}

The topological constraints on Brownian walks are a long standing subject
of interest in mathematics and polymer physics. A particularly fruitful
 example is the study of  the winding angle $\theta$, the continuous
 angle swept by the movement
 around a set of points (curves) in two (three)
 dimensions  \foot{Since all results are even
in $\theta$ in what follows we generally
consider $\theta\geq 0$. Otherwise, substitute $|\theta|$ for $\theta$}.
 In a plane, the probability distribution at large time or length
 $l$ for the winding angle of a Brownian walk  around any point O
 was determined by Spitzer \ref\S{F.Spitzer, Am. Math. Soc.
 87 (1958) 187.} as
\eqn\spi{P\left[x={2\theta \over\ln l}\right]={1\over\pi}{1\over 1+x^2},\
l\rightarrow\infty.}
In  what follows the length of the walk is  usually thought of as the
 number of steps in  a discretized version. Because of the logarithmic
dependence on $l$ the various possible definitions of $l$, all proportional
 to each other, are equivalent. This of course is compatible with angles being
dimensionless.

The law \spi\ is substantially modified when a cutoff is introduced so that
 the walk cannot get arbitrarily close to O. One finds then \ref\RH{J.Rudnick,
 Y.Hu, J.Phys. A20 (1987) 4421.}
\eqn\cuto{P\left[x={2\theta\over\ln l}\right]={\pi\over4}
{1\over[\cosh\pi x/2]^2},\
 l\rightarrow\infty .}
The standards methods of approach to these problems are based
 on diffusion equations, functional integrals or refined
 probability
 theory \ref\E{S.F.Edwards, Proc. Phys. Soc. London 91
 (1967) 513.}\ref\PY{J.W.Pitman, M.Yor, Ann. Proba. 14 (1986) 733.}
\ref\CDB{A.Comtet, J.Desbois, C.Monthus, preprint IPNO/TH 93.}.

More recently the same problem has been considered for self
 avoiding walks (SAW). By applying methods derived from Coulomb Gas
representations and conformal field theory ,
 the equivalent of \spi\ or \cuto\ (for SAW
 there is  no difference between the
two cases  as the walk provides a natural UV cutoff, see figure 1)
 was determined \ref\DS{B.Duplantier, H.Saleur, Phys.
 Rev. Lett. 60 (1988) 2343.}
\eqn\saw{P\left[x={\theta\over (4\ln l)^{1/2}}\right]={e^{-x^2}
\over\sqrt{\pi}},\ l\rightarrow\infty.}

The purpose of this note is to study the relation between \cuto\ , \spi\
 and \saw\ further, and to compute a related distribution for SAW
that will also depend on the variable  $\theta/\log l$. The analysis
 uses corner transfer matrix ideas and conformal field theory. It can be
 considered as a follow up of \DS\ .

 Indeed, the most striking  difference between the Brownian and
self avoiding cases
 lies in the variable that is either $\theta/\log l$ or $\theta/(\log
l)^{1/2}$.
 The intuitive explanation of this difference  is that the distribution
in the self avoiding case is mainly determined by the excluded volume of
 the walk already wound around O (figure 1), while in the brownian case,
there is no
such effect, and the distribution is determined fully by the entropy loss
 that arises from the constraint of winding .
To suppress this major difference and
concentrate on curvature related entropy we
 simply put  the SAW on the multi sheeted Riemann surface for the
function $\ln z$,
 discretized if necessary. For a walk of length $l$
 we define the probability of winding angle $\theta$ by
the relative number of configurations that sweep a total
 angle $\theta$, and for $|\theta|>2\pi$ have end points
 on different sheets. It is important
 to realize  that this problem would be {\bf identical} to the
 usual winding angle problem for the brownian case (since there is no
interaction
between different parts of the walk one can just collapse the staircase
onto the plane).

 To determine the probability distribution we can  proceed in two ways.

\newsec{Conformal mapping}

{\bf 1.} First we consider a lattice SAW on a strip of length L and width W.
The boundary conditions are periodic (free) in the L (W) direction (figure 2).
 On an
infinite lattice the number of SAW of length l grows as $\Omega_l\propto\mu^l$,
 where $\mu$ is the lattice effective coordination number. On the strip,
the number of configurations for  SAW winding around the periodic strip is
 denoted by $\Omega^{L,W}_l$. We then consider the generating function
\eqn\zz{{\cal Z}_1=\sum_l\Omega^{L,W}_l\mu^{-l},}
 In the limit where $L/a,W/a\rightarrow\infty$, $L/W$ remaining finite
 (a is the lattice spacing),
$Z$ is determined by conformal invariance and Coulomb gas mappings arguments.
 It is important to notice that there is {\bf no bulk term},
the reason being that SAW have a fractal dimension less that
 two. The rest of the procedure is standard, but we recall
 it very briefly for completness. First, we choose for technical
 reasons the honeycomb lattice, and reformulate the polymer problem
as the limit $n\rightarrow 0$ of an $O(n)$ model. Partition and
 correlation functions of this $O(n)$ model are expressed as sums
 over self avoiding mutually avoiding sets of loops and open walks,
 with a weight $\beta$ (the inverse temperature) per monomer and $n$ per
 loop \ref\N{B.Nienhuis, Phys. Rev. Lett. 49 (1982) 1062.}. For $n=0$,
 the critical temperature is $\beta_c=\mu^{-1}$. Next, one reformulates
 the $O(n)$ model in terms  of the Izergin Korepin (IK) vertex model
\ref\IK{A.G.Izergin, V.Korepin, Comm. Math. Phys. 79 (1981) 303.}
\ref\NI{S.O.Warnaar, M.T.Batchelor, B.Nienhuis, J.Phys. A25 (1992) 3077.},
 the correspondence between the two being conveniently carried out using
 quantum group symmetries \ref\PS{V.Pasquier, H.Saleur, Nucl. Phys. B330
 (1990) 523.}\ref\FSZ{P.Fendley, H.Saleur, Al.B.Zamolodchikov,  preprint
USC-93-003.}. The continuum limit of the polymer problem is then
 worked out using a combination of sectors of the continuum limit of the
 IK model, which is described by a Gaussian model with charge at infinity
\ref\DF{V.Dotsenko, V.A.Fateev, Nucl. Phys. B240 (1984) 312.}. The net
 result for the lattice model is the following. The partition function
 where all non contractible loops on the cylinder have weight $n=0$ has
 the expression
\eqn\zI{{\cal Z}_0=-\sum_{p=0}^{\infty}d_p{\cal K}_p+d_{p+1/2}
{\cal K}_{p+1/2},}
where ${\cal K}_s$ is the partition function of the IK model
in the spin $s$ (integer or half integer) sector, and $d_s$
 is the quantum dimension
\eqn\dd{d_s={t^{2s+1}-t^{-2s-1}\over t-t^{-1}},}
with
\eqn\nn{n=-t-t^{-1}.}
At $t=i$, using the additional level coincidences induced by
 $U_tsl(2)$ representation theory at $q$ a root of unity \PS\ , one
checks
that ${\cal Z}_0=1$ (see \ref\BS{M.Bauer, H.Saleur, Nucl. Phys. B320
 (1989) 591.}). The computation of ${\cal Z}_1$ follows from the
correspondence between the IK model  and  polymers \FSZ\ \BS\ as
\eqn\zII{{\cal Z}_1={\partial{\cal Z}_0\over \partial n}
-{1\over 2}{\partial\over \partial t}
{\cal Z}_0=\sum_{p=0}^\infty (p+1)(-1)^p{\cal K}_{p+1/2}.}
In the continuum limit ($L/a,W/a\rightarrow\infty, L/W$ finite) we use
the standard result that \BS\
\eqn\kk{{\cal K}_s\rightarrow {q^{h_{1+2s,1}}-q^{h_{-1-2s,1}}\over P(q)},}
where
\eqn\qq{q=\exp\left(-{\pi L\over W}\right)}
and  $P(q)=\prod_{n>0}(1-q^n)$, and $h_{r,s}$ denotes the
 standard conformal weights in the Kac notations \ref\BPZ{A.A.Belavin,
A.M.Polyakov, A.B.Zamolodchikov, Nucl. Phys. B241 (1984) 333.},
 with, in the polymer case, $h_{r,1}={(3r-2)^2-1\over 24}$. We obtain then
\eqn\ZZ{{\cal Z}_1\rightarrow Z_1=-{1\over P(q)}\sum_{n=-\infty}^
\infty n(-1)^nq^{(6n-1)(2n-1)/8},}
One can also  consider the same quantity $Z_2$ defined for two
 SAW, mutually avoiding, that wrap around the cylinder. One has
\eqn\ZZI{Z_2=-{1\over P(q)}\sum_{n=-\infty}^\infty n(n+1)(-1)^n
q^{3n(3n+1)/6}.}

\bigskip

{\bf 2.} In the limit $L,W\rightarrow\infty$ the lattice strip
 can be considered as continuous, and  $Z$ is a partition
 function for the conformal field theory that describes
polymers. The various results regarding this conformal field
 theory can be followed through conformal mappings. We consider
 the mapping $z'/a=\exp(z/a)$ where $z$ is the complex variable
in the strip plane, the time direction $L$ being along
 the purely imaginary axis. We then obtain the same generating
 function but  for SAW wrapping on a staircase (with the two end
"lips" identified, see figure 3) of inner radius
 r and outer radius R in ratio $W/a=\log (R/r)$,
 and angle  $\theta= L/a$:
\eqn\qqI{q=\exp\left[-{\pi\theta\over\ln(R/r)}\right].}
 To show this  more precisely it is necessary to identify
$Z$ with the trace of a transfer matrix, and then
 use known results for  eigenvalues, whose properties
 in the mapping are deduced from the ones of correlation
functions of the theory \ref\C{J.Cardy, J.Phys. A17 (1984) L385.} .
 We explain this in more details in the next section.

 It is important to realize the exact meaning of the angle $\theta$:
 our results make sense when $r/a,R/a\rightarrow\infty, R/r$ finite {\bf and }
 $\theta\rightarrow\infty,\theta/W$ finite, since on the strip we had
 $L/a\rightarrow\infty$.
The above result still holds when $r$ remains of the
 order of the lattice spacing, while $R,\theta$ still are very
 large \BS\ \ref\PT{I.Peschel, T.T.Truong, Z.Phys. B 69 (1987)
 385.}\ref\CP{J.Cardy, I.Peschel, Nucl. Phys. B300 (1988) 377.}.
This is due to the logarithmic dependence of $W$ on the radiuses ratio:
we discuss  a similar case in the next paragraph. In the following we
 take $r=a$.

\bigskip

{\bf 3.} Third we must discuss what becomes of
this generating function when evaluated in the
 scaling region  $\beta\approx\mu^{-1}$ (in fact we will exclusively deal
with the high temperature part $\beta<\mu^{-1}$. The behaviour for
low temperatures is quite different because of symmetry breaking
\ref\FS{P.Fendley,
H.Saleur, Nucl. Phys. B388 (1992) 609.}).
 Once again we can
 rely on the $O(n)$ model analysis. Recall that the limit
 $\beta\rightarrow\mu^{-1}$ is the approach to the critical point,
 where the correlation length diverges as
\eqn\kkki{{\xi\over a}\propto(\mu^{-1}-\beta)^{-\nu},}
 where $\nu$ is also the  exponent characterizing the size of the SAW
($<(R_G/ a)^2>\propto l^{2\nu}$, $R_G$ the radius of gyration).

Now, in most problems, the knowledge of some quantity at
the critical point is not enough to determine this quantity
away from it. Usually the functional dependence is constrained
 by scaling arguments, but a non trivial function has still to
 be determined. The present problem turns out to be  simpler.
Indeed the usual finite size scaling hypothesis, valid in the
limit $R/a,\xi/a\rightarrow\infty$, involves replacing for
 each of the terms in the above sums $R$ by $Rf(\xi/R)$ where
the function $f$ depends on the term considered. One needs
 $f(\infty)=1$ and also, to suppress the $R$ dependence
in case the physics is dominated by the much smaller scale $\xi$,
 $f(x)\approx f_0 x,\ x\rightarrow 0$ \ref\DG{P.G.De Gennes,
"Scaling concepts in polymer physics", Cornell University Press.}.
 Now suppose we consider precisely the latter regime.
 Since $\xi$ still has to be very large compared to
 the lattice spacing and the dependence is logarithmic,
 $f_0$ disappears from \ZZ\ \ZZI\  as $\log (f_0 \xi/a)\approx
\log (\xi/a), \xi/a\rightarrow\infty$. Therefore we conclude
that in the scaling region, \ZZ\ \ZZI\ still hold with $R/a$
replaced by $\xi/a$:
\eqn\qqII{q=\exp\left[-{\pi\theta\over\ln(\xi/a)}\right].}
 In effect, the IR cutoff at criticality is exactly replaced by
 the correlation length in the scaling region.

\bigskip

{\bf 4.} So far we have discussed   generating functions
only, while results of most statistical interest rather deal with
 SAW of large but  fixed length. The standard procedure
 to get such results involves inverse Laplace transform \DG\ ,
 and is usually quite complicated for  full probability distributions.
 Here again, the situation is simplified by the logarithmic dependence.
Call $\Omega^\theta_l$ the number of SAW of length $l$ on the staircase.
 The scaling function $F$ we are looking for is defined by
\eqn\sc{\Omega^\theta_l\approx \mu^ll^{-1}(\ln l)^\alpha F\left({\theta\over
 \ln l}\right).}
The form of \sc\ is explained as follows. For $\theta$ small
 we should recover the usual result for a self avoiding loop,
taking also into account the combinatorial fact that our loop
can assume any position provided it encircles the origin \DG\ .
 This determines that the power $l^{-1}$ in $\Omega^\theta_l$.
 The exponent $\alpha$ is not known; it does not affect our results,
and we set it to zero in the following. The generating function we are
 computing is then
\eqn\gg{{\cal Z}^{plane}_1=\sum_l\Omega^\theta_l\beta^l\approx
\int F\left({\theta\over\ln l}\right)\exp\left[l(\beta\mu-1)
\right]{ dl\over l}.}
 Since in the  region  of integration $l>>1$, $F$ in \gg\
varies very slowly over the domain where the other term in
the integrand takes appreciable values. This domain is around $l\propto
{1\over 1-\beta\mu}$. Since $F$ depends logarithmically on $l$ and we
 consider the behaviour as $\beta\rightarrow\mu^{-1}$, we conclude that
 up to proportionnality constants ${\cal Z}_1^{plane}\propto F\left[-
{\theta\over\ln (1-\beta\mu)}\right]$ ie  the fixed length distribution $F$ is
 obtained by replacing in ${\cal Z}_1^{plane}$ ( \ZZ\ or \ZZI\ )
$\ln(\xi/a)$ by $\ln l^\nu$. In other words, the IR cutoff is now
 given by the average linear size of the chain.We find thus
\eqn\FF{F\left(\theta\over \ln l\right)\propto-{1\over P(q)}
\sum_{n=-\infty}^\infty n(-1)^n q^{(6n-1)(2n-1)/8},}
where
\eqn\qq{q=\exp\left(-{\pi\theta\over\nu\ln l}\right).}
The distribution function $F$ has  the following asymptotic behaviours
\eqn\asy{F(x)\propto \exp(-5\pi x/8),x\rightarrow\infty;\ F\propto
{1\over x},x\rightarrow 0.}
where the last result follows from Poisson resummation. The
 divergence at small argument arises from the fact that in our
computation the origin of the SAW is not fixed, introducing an
additional degree of freedom. See section 4.

\bigskip

\newsec{Corner transfer matrix}

We can also derive this distribution using corner transfer matrix
 ideas, which are a little hidden above.

 We now work  in the plane from the start. We  can still use
the polymers,  $O(n)$, IK model correspondence. The main difference
with the computation on the strip is that the evolution is described
by a corner transfer matrix (CTM) instead of a row to row  one. It is
 important to notice that the CTM has also the property of commuting
with $U_tsl(2)$: this is most clearly seen in the hamiltonian limit
where for the row to row transfer matrix, $H=\sum H_i$, and for the
 CTM, $K=\sum iH_i$ \ref\T{H.B.Thacker, Physica 18D (1986) 348.}
where $[H_i, U_tsl(2)]=0$. As a result, the generating functions
 ${\cal Z}^{plane}$ have  expressions similar to the ones on the
 cylinder, like eg
\eqn\zzp{{\cal Z}_1^{plane}=\sum_{p=0}^\infty (p+1)(-)^p{\cal L}_{p+1/2},}
where now ${\cal L}_{s}$ is the partition function of the IK model
 in the "radial spin $s$ sector" \ref\B{R.J.Baxter, "Exactly solved
 models in statistical mechanics", Academic Press.},
 ${\cal L}=Tr\left(\tau_{CTM}\right)^\theta$. Similar expressions
 appear in the computation of spontaneous magnetization for the
 $Q>4$ state Potts model \ref\BI{R.J.Baxter, J.Phys. A15 (1982) 3329.}.
 In corner transfer matrix calculations $\theta$ is usually a
multiple of $\pi/2$, but in the large angle limit we consider
 this becomes irrelevant.

Now the problem is to evaluate directly (without using the chain
of arguments of the previous section) the continuum limit of
 ${\cal L}$ when $\beta\rightarrow\mu^{-1}, \theta\rightarrow
\infty$. When $\xi/a=\infty $ it is known that the IK model
 corresponds to a free bosonic  theory with action \N\
\eqn\sg{A=\int -{g\over 4\pi}(\partial\phi)^2 ,}
where for polymers $g=3/2$, $\phi$ is the usual free field variable,
 dual to the arrows of the IK model in its solid on solid reformulation
 \N\ with normalization $\delta\phi=2\pi\times\hbox{ spin}$. Now when
 $\xi/a>>1$ this action has to be modified by adding suitable terms
 that make the model massive. Suppose first we were dealing with a
free massive boson, ie $\delta A=m^2\int \phi^2$, $m=1/\xi$. Suppose
 we consider a sector of radial spin $s$ ie with boundary conditions
\eqn\bc{\phi(r=\infty)-\phi(r=a)=-2\pi s.}
The classical solution (ie satisfying $(\Delta +m^2)\phi_{cl}=0$)
 to this equation is
\eqn\phicl{\phi_{cl}=2\pi s{K_0(mr)\over K_0(ma)},}
where  $K_0,K_1$ are the usual  Bessel functions. One has
\eqn\terms{(\partial\phi_{cl})^2=4\pi^2s^2m^2\left[{K_1(mr)\over
 K_0(ma)}\right]^2,}
and
\eqn\termsI{m^2\phi_{cl}^2=4\pi^2s^2m^2\left[{K_0(mr)\over K_0(ma)}\right]^2.}
We now have to evaluate the contributions of these two terms to
the action, ie compute $\theta\int rdr$  in the limit where $ma<<1$. Using
\eqn\asym{K_0(x)\propto\ln(x),\ K_1(x)\propto{1\over  x},\ x\rightarrow 0,}
it is easy to see that the dominant contribution comes from
the {\bf kinetic term}, behaving as $1/\ln (ma)$ while the mass
term gives a contribution of  order $1/\ln^2(ma)$.The classical
 part therefore behaves as in the analysis of the critical model
 on the periodic strip, with the now familiar substitution $L/W\rightarrow
\theta/\ln(1/ma)$.  Therefore we conclude that in the  limit
 $ma\rightarrow 0,\ \theta\rightarrow\infty, \theta/\ln(ma)\hbox{ finite}$
\eqn\scl{{\cal L}_{s}\rightarrow Z_{qu}\left(q^{(6s+1)^2/24}-
q^{(6s+5)^2/24}\right).}
Now we must determine $Z_{qu}$. This is most easily done by studying
 the diffusion equation for Brownian motion on our Riemann surface
(not surprisingly,  an identical problem occurs in the study of the
 winding angle distribution for Brownian walks \RH\ ). Calling $G$
 the Green function one finds (recall that the region of radius $a$
 around the origin is excluded in our problem)\RH\
\eqn\green{G(r_<,r_>,\theta)=\int_{-\infty}^\infty e^{i\nu\theta}K_\nu(mr_>)
\left[I_\nu(mr_<)K_\nu(ma)-I_\nu(ma)K_\nu(mr_<)\over K_\nu(ma)\right]\ d\nu}
Using contour integration, this can be rewritten as an infinite discrete
 sum over the zeroes (in variable $\nu$) of the denominator, corresponding
 to a sum over eigenvalues of $\tau_{CTM}$. In the limit $ma\rightarrow 0$
 these zeroes occur at
\eqn\zeroes{\nu_n\approx {in\pi\over\ln(1/ma)}}
contributing to the angular dependence of the Green function by a
term $\exp[\pi\theta n/\ln(ma)]$. We have therefore determined
the eigenvalues of the CTM, up to degeneracies. We now can rely
on a mode analysis and the study of the ground state energy in
 \CP\ to conclude that
\eqn\res{Z_{qu}={q^{-1/24}\over P(q)},}
where $q=\exp[-\pi\theta/\ln(1/ma)]$. Although  the correct
 action describing polymers in the scaling region is rather
the sine-Gordon model, it is easy to see that in the limit
we are interested in, this does not make any difference:
 all that matters is the kinetic term and the presence of
 a length scale $m=1/\xi$. Therefore combining \scl\ and
\res\ we recover the results of the previous section.

The careful reader may notice that in usual corner transfer matrix
 computations
there are additional terms, usually made of elliptic functions, that
multiply the exponential of the corner transfer matrix hamiltonian \B\ .
 In fact  they disappear in the scaling limit we study here.
Finally, computations on an integrable version of the off critical polymer
problem have recently appeared, which confirm our results
\ref\NW{D.Nemeschansky,
N.P.Warner, preprint USC-93-023.}.

\newsec{Comments}

{\bf 1.} The distribution \cuto\ holds for a problem where the origin and
 end points of the Brownian walk need not be on top of each other. It
 looks quite difficult to obtain a similar quantity for SAW; instead
 of taking a trace of the transfer matrix, one also has to insert matrix
 elements of boundary operators, which are known in principle but quite
 complicated. It is easier to derive the winding angle distribution for
 Brownian loops. One finds
\eqn\bl{P\left[x={2\theta\over\ln l}\right]\propto{1\over [\sinh\pi x/2]^2},}
which diverges at small argument, like \asy\ did.

\bigskip

{\bf 2.} The distributions \bl\ and \cuto\ for Brownian
 walks and \FF\ for SAW  are similar in nature. To see
this more explicitely expand \cuto\ as
\eqn\cutoI{P\left[x={2\theta\over\ln l}\right]=\pi\exp(-\pi x)\left(\sum_{n=0}
^\infty (-)^n\exp(-\pi nx)\right)^2=\pi q_B\left(\sum_{n=0}^\infty(-)^nq_B^n
\right)^2,}
where
\eqn\qb{q_B=\exp\left(-{\pi\theta\over {1\over 2}\ln l}\right),}
an expression similar to \qqII\  since the exponent $\nu$ for
Brownian walks is equal to $1/2$.

 The UV cutoff
necessary to get \cuto\ and not \spi\ appears now very natural from the
 corner transfer matrix or conformal mapping point of view.
 Distribution \spi\ arises when this cutoff disappears and the CTM
spectrum is not quantized any more: this situation is not met
in most other lattice  models (like the  8 vertex model) where the spectrum
of the CTM is discrete, and in the continuum there is always an implicit
UV cutoff like in the SAW case.

Let us  emphasize that the integer spaced structure of the CTM spectrum
{\bf away} from the scaling region in integrable models \B\ has nothing to
do with the above arguments. We also do not believe that this integer spaced
structure
is related to quantization on angular momentum as claimed in \T\ : this is
especially
clear in the scaling region.

\bigskip

{\bf 3.} The same procedure could also be applied to revover
the result of \DS\ concerning a SAW on the plane that makes
 many turns, building in that fashion a core of forbidden region aroud O
 (figure 1). In this latter case  one  deals indeed  with the plane,
 not the  infinite staircase so
 $L/W\rightarrow 0$ and the winding angle appears in the spin
 variable $s$. Since the dimension of the ground state of the
 spin $s$ sector grows like  $s^2$, this explains why the
 distribution \saw\  depends on $\theta^2/\ln l$. Let us
take this opportunity to complete a point that was overlooked
in \DS\ . The scaling form of a row to row transfer matrix
 eigenvalue as deduced from conformal invariance
$\Lambda\approx\exp(-\pi h/W)$ holds  only in the limit
 $W\rightarrow\infty$ with $s$ {\bf finite}. To establish
the Gaussian distribution of \DS\ one needs to apply the
same formula for $s\rightarrow\infty$ too, but $s\propto\sqrt{W}$ so
 $s<<W$ still. We do not know any proof that this is legitimate, but
 we can give arguments based on the correction to scaling analysis
 \ref\CI{J.Cardy, Nucl. Phys. B270 (1986) 186.}. Considering for
instance the analytic corrections induced by the terms $L_{-2}\bar{L}_{-2}$
and $L_{-2}^2,\ \bar{L}_{-2}^2$ (where $L_{-n}$
denotes the usual Virasoro algebra generators \BPZ\ ) one finds that the
scaled log of eigenvalue
 behaves as $s^2+c{s^4\over W^2}$ so even as $s\propto\sqrt{W}$
 the result still holds.

\bigskip

{\bf 4.} In conclusion, the winding angle distributions
of the Brownian, SAW and probably other geometrical problems (like percolation)
 around a single point can be deduced from conformal field
theory only, and are given by some kind of theta function.
Notice that such distributions  make reasonable sense as
statistical problems only because the bulk free energies vanish.
Otherwise the results would always be dominated by very large angles. We do not
really believe that our distribution \FF\ on a staircase has much practical
interest. However it is amusing to notice that it can be reformulated
slightly differently. "Unfold" the discretized periodic staircase  on the plane
as in figure 4. Locally this is like the plane, but there is curvature
at the origin. Since in our SAW problem all edges are treated as having the
same length, we consider this lattice as the triangulation of a manifold
with curvature located at the vertices, $\rho_i=\pi{4-q_i\over q_i}$
 where $q_i$
is the number of neighbours of vertex i (with the surface element
 $s_i={q_i\over 4}$
this satisfies Gauss Bonnet theorem for closed manifolds). Distribution \FF\
appears thus as the distribution for a SAW "trapped" around a point of strong
 negative curvature.
in the limit where $l,\rho\rightarrow\infty$. Maybe in that form will it prove
more useful.

\bigskip
\noindent {\bf Acknowledgments}: I thank J.Rudnick and N.P.Warner for useful
discussions. This work was supported by the Packard foundation.

\vfill\eject
\listrefs
\centerline{\bf Figure Captions}
\bigskip

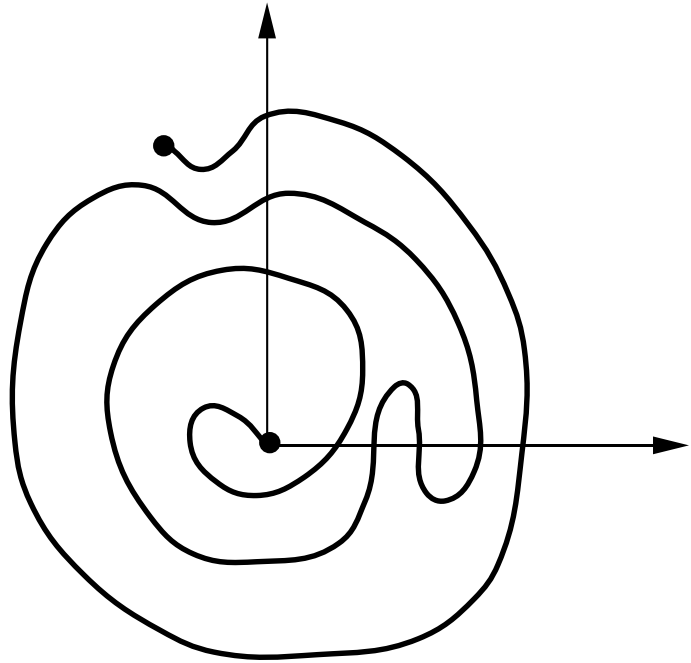
\noindent Figure 1: A SAW on the plane winding around the origin.

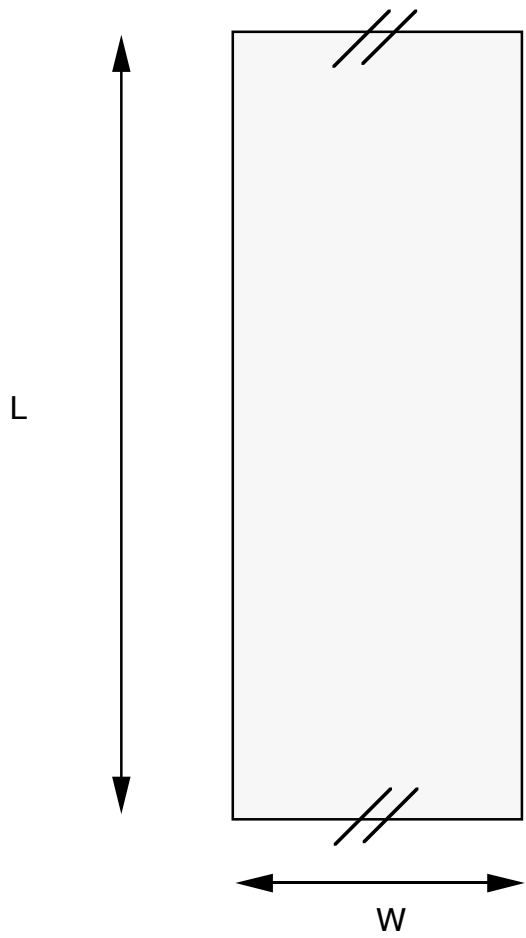
\noindent Figure 2: Periodic strip.

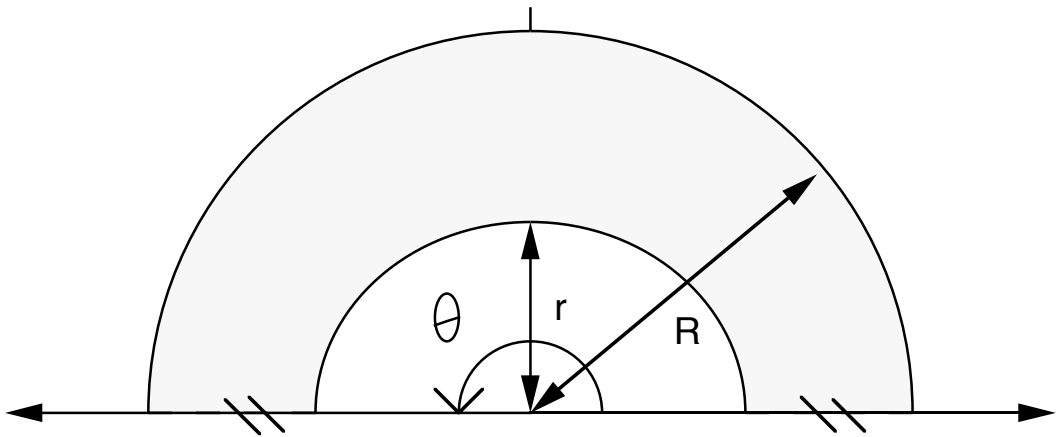
\noindent Figure 3: By conformal mapping the strip maps onto a periodic
staircase.

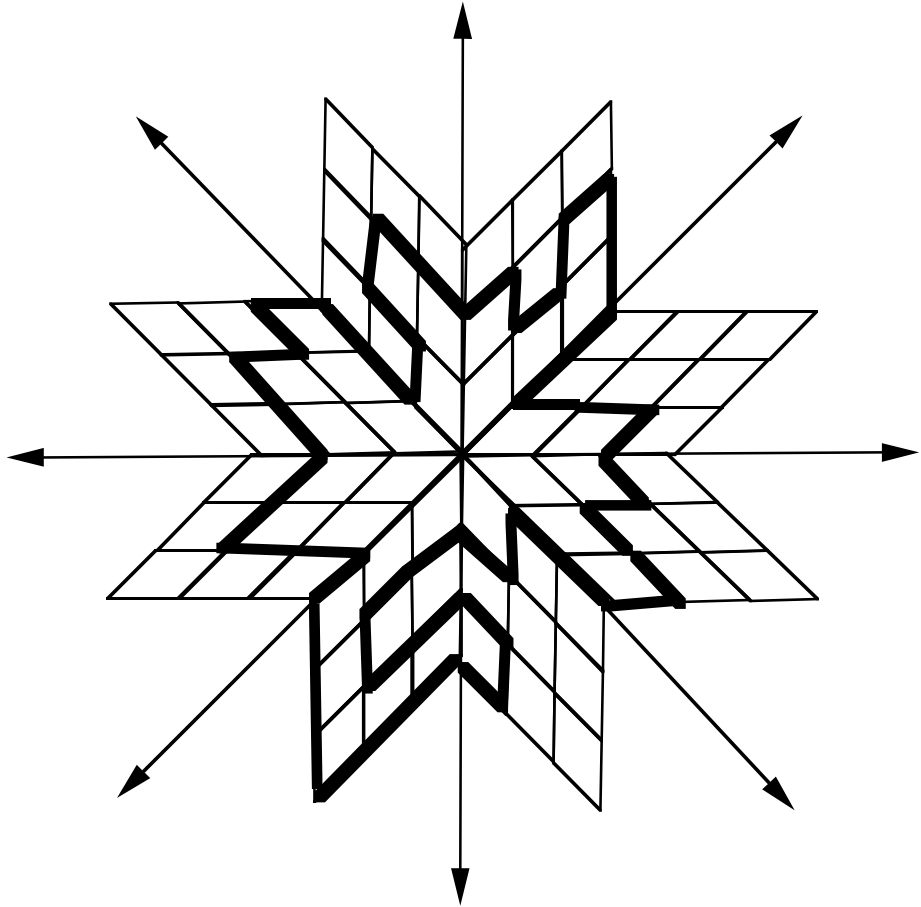
\noindent Figure 4: A discretized staircase "unfolded" on the plane with
negative
curvature located at the origin, and a self avoiding loop that winds around it
(with $\theta=8{\pi\over 2}=4\pi$).
\bye